\documentclass[letterpaper,11pt,conference]{IEEEtran}
\usepackage{tikz}
\usepackage{listings}
\usepackage{cite}
\usepackage{rotating}
\usepackage{multirow}
\usepackage{array}
\usepackage{caption}
\usepackage{dblfloatfix}

\definecolor{olivegreen}{rgb}{0.2,0.8,0.5}
\definecolor{grey}{rgb}{0.5,0.5,0.5}
\lstdefinelanguage{ttl}{
sensitive=true,
tabsize=4,
morecomment=[l][\color{grey}]{@},
morecomment=[l][\color{olivegreen}]{\#},
morestring=[b][\color{blue}]\",
}

\usetikzlibrary{positioning}
\usetikzlibrary{arrows}

\tikzset{entity/.style={rectangle,draw,rounded corners,align=center,minimum height=3em}}
\tikzset{activity/.style={rectangle,draw,align=center}}
\tikzset{attribute/.style={rectangle,draw,align=center}}
\tikzset{conn/.style={arrows={-triangle 45}}}
\tikzset{connlabel/.style={font=\ttfamily,text=,scale=0.7}}

\begin{document}
\nocite{*}
\title{Enhancing Information Awareness Through Directed Qualification of Semantic Relevancy Scoring Operations}

\author{\IEEEauthorblockN{Jason~Bryant, Greg~Hasseler, Matthew~Paulini, Timothy~Lebo} \IEEEauthorblockA{Air Force Research Laboratory - RISA Information Management Technologies \\
Distribution A: Approved for public release; distribution is unlimited. \\
(Approval AFRL PA \#88ABW-2014-1863)}}

\newpage
\onecolumn
\makeatletter
\begin{center}
\textbf{19th ICCRTS} \\

\vspace{10 mm}

\textbf{Title of Paper:} \@title \\

\vspace{10 mm}

\textbf{Assigned Paper \#:} 027 \\

\vspace{10 mm}

\textbf{Primary Topic:} Data, Information and Knowledge \\

\vspace{10 mm}

\textbf{Alternatives:} Organizational Concepts and Approaches; Experimentation, Metrics, and Analysis \\

\vspace{10 mm}

\textbf{Authors:}

Jason~Bryant, Greg~Hasseler, Matthew~Paulini, Timothy~Lebo \\

\vspace{10 mm}

\textbf{Organization:} \\
Air Force Research Laboratory \\
RISA - Information Management Technologies \\
525 Brooks Road \\
Rome, NY 13440 \\

\vspace{10 mm}

\textbf{Point of Contact:} Jason Bryant \\
\textbf{Organization Phone:} (315)330-7670 \\
\textbf{Organization Email:} Jason.Bryant.8@us.af.mil \\

\vspace{10 mm}

Distribution A: Approved for public release; distribution is unlimited. \\
(Approval AFRL PA \#88ABW-2014-1863)

\end{center}\makeatother
\twocolumn
\newpage

\maketitle

\begin{abstract}
Successfully managing analytics-based semantic relationships and their provenance enables determinations of document importance and priority, furthering capabilities for machine-based relevancy scoring operations. Semantic technologies are well suited for modeling explicit and fully qualified relationships but struggle with modeling relationships that are qualified in nature, or resultant from applied analytics. Our work seeks to implement the autonomous Directed Qualification of analytic-based relationships by pairing the Prov-O Ontology (W3C Recommendation) with a relevancy ontology supporting analytics terminology. This work results in the capability for any semantically referenced document, concept, or named graph to be associated with the results of applied analytics as Direct Qualification (DQ) modeled relational nodes. This new capability will enable role, identity, or any other content-based measures of relevancy and analytics-based metrics for semantically described documents.
\end{abstract}

\section{Introduction}
Robustly supporting mission solutions within a C2 information domain is difficult when mission information needs cannot be satisfied with currently available information, fluctuate, or if there are unforeseen state changes for mission circumstances. C2 Agility defines how to flexibly support operations within a single C2 space, as well as how and when to appropriately transition between C2 Approaches for a particular mission and its circumstances. Ideally, maximizing agility would provide tailored solutions for adjusting to changes that occur within the circumstances of the mission. In practice, however, changing mission circumstances can be managed more simply through enhanced mission modeling, while a lack of suitable information that assists the deconfliction of prioritization-based mission states is more difficult to solve.

The lack of suitable information, or which lack relevance measures for determining suitability, is a relatively common challenge, yet is resist to mission modeling solutions. Our approach seems to provide mission operations with information that was previously non-existent, non-queryable, or incorrectly deemed irrelevant. Advanced features of semantic mission modeling can be used in concert with traditional approaches to increase agility in the face of insufficient information. This results in meeting mission information needs by increasing the quality of information available by determining relevancy, and increasing the quantity of information through mission domain semantics.

The primary focus of this effort is to enable new features for semantics-based relevancy modeling and analytics. Semantic-based graphs traditionally only support reasoning that is logic-based, whereas non-semantic knowledge graphs are commonly evaluated using traversal queries, popularity, similarity, clustering, or other analytics. Analytics-based graph analytics have never been sufficiently modeled ontologically in order to enable being paired with semantic inferencing engines. Direct Qualification (DQ) is the means by which semantic modeling can be applied to express these probabilistic or analytics-based relationships and bridge the current divide between semantic inferencing logic and graph-based analytics.

In this paper, we show that Directed Qualification, introduced by the Prov-O ontology, can be combined with a relevancy ontology for modeling analytical evidence supporting semantic relationships while avoiding the pitfalls of reification. This enables techniques used to evaluate non-semantic knowledge graphs to be applied to semantic-based knowledge graphs. We demonstrate this ability by applying the Betweenness, PageRank, and HITS analytics to a semantic-based knowledge graph derived from an operational scenario.

While there are ongoing efforts towards document analysis using analytics such as PageRank ~\cite{Ding_2002_PageRank}, VSM, HITS, etc., the focus of these efforts has been on ontology matching ~\cite{Tous_2006_Vector} or temporal/geospatial query enhancement ~\cite{Perry_2009_Geospatial}. Our approach differs in that it stays confined to semantic technologies. Enabling analytic provenance relationships and the qualification of applied analytics for interrelated documents provides critical capabilities for result set relevancy scoring, however, some intrinsic difficulties within semantic standards and technologies require mitigation.

In the following section, we discuss why the ability to model analytical evidence in semantic-based knowledge graphs is valuable to Command and Control (C2) environments. Section \ref{sec:modeling} describes the technical work necessary to implement directed qualification for modeling analytics. Section \ref{sec:experiment} describes our experimental setup, while Section \ref{sec:relevancy} describes the results. Finally, Section \ref{sec:conclusion} concludes our work and discusses potential applications and future work.

\section{Background}
\label{sec:background}
A current complexity in C2 support is adequately defining the transition requirements triggering a change in the C2 Approach for a set of entities or collectives. The triggers are generally event-based from either information publications within the mission's C2 domain, or a set of entity interaction pattern occurrences from across one or more C2 domains (e.g., Edge, Conflicted, etc.). Our approach focuses on the evaluation and semantic modeling of event-based information publications springing from mission resources. We demonstrate how greater command and control agility is achieved through pairing document-based semantic relevancy scoring and modeling with graph-based analytics and traceability to persisted source documents.

Provenance for semantically represented documents provides support for attribution of authorship, change tracking, sourcing, and any other transform related to an entity, activity, or agent involved with the document. Our research uses the Prov-O[REF] Ontology and direct qualification of semantic relevancy scores to manage these relationships.

Semantic provenance for documents supports a slightly different subset of features than semantic provenance for non-document-centric relationships. Non-document-centric relationships lack the context of sourcing unless explicitly and independently associated via provenance relationships. The consequent semantically expressed relationships lack traceability.

Semantic provenance for documents is handled less explicitly by using the unique document URI as the named graph, or quads, for all relationships extracted from the document. This traceability is important because without introspection into referential sourcing there is a limited measure of trust in data content. For non-deterministic relationships, the implications of using semantics grow even more complex. Many probabilistic analytics improve reliability and trust when trends of an increasing or decreasing score are tracked over time.

A key advantage of pairing provenance and relevancy vocabularies is the power it provides to overcome the complexity of semantic reification. Reification, an intrinsic complexity of the semantic standards, is the consequence of attempting to simplify all relationships into Subject-Predicate-Object sets. It is normally implemented when a semantically modeled instance is seeking to express either the qualification or provenance of a relationship. These two cases can be mitigated without resorting to reification, however. Adopting a quad-based perspective of semantic relationships can achieve a basic form of provenance by allowing traceability to the source named graph's unique URI. The Prov-O ontology expands the set of provenance support and supplies some generalized predicates for qualification. This effectively solves the non-probabilistic subset of analytics use cases. However, even with pairing both of these approaches there is a failure to solve the qualification of probabilistic analytics, such as the results of Vector Space Modeling, PageRank, HITS, or other Natural Language Processing (NLP) analytics. Our approach towards supplying these capabilities is the Direct Qualification of probabilistic relationships with a supporting relevancy ontology.

The primary steps for enabling Direct Qualification are outlined below, and described more completely in Sections \ref{sec:modeling}, \ref{sec:experiment}, and \ref{sec:relevancy}.
\begin{enumerate}
	\item Support persistence for raw documents and semantic quad-based relationships, ideally by using the semantic URI of the document's named graph as the unique key for the raw document retrieval.

 	\item Strictly enforce the separation of the semantic models for class instances from the events affecting their state relationships.

 	\item Support graph-based processing of analytics over semantic edges and vertexes.

	\item Support event-based relevance scoring triggers, such as SPARQL queries, XPath queries, semantic reasoning, or keyword searches of raw text.

	\item Determine the appropriate Direct Qualification Model based upon tests for occurrence, continuance, and the monotonicity of the entities involved in the applied analytic.

	\item Express the relevancy scoring of documents through the pairing of a provenance ontology with a relevancy ontology.

	\item Persist the DQ results within the quad-store.
\end{enumerate}

\section{Motivation And Related Work}
\label{sec:motivation}
Data can be looked at from multiple domain-agnostic viewpoints, although each possesses intrinsic advantages and disadvantages. For example, the Internet can be viewed as a series of interconnected documents, or, if all content were to be extracted, as a set of interconnected facts, social networks, people, geo-locations, media, and applications. While these views are not mutually exclusive, applications are generally oriented towards a single paradigm. C2 environments share a great deal of similarities with these paradigms, even if obscured by segmented tiers of access control, middleware management tools, multiple security configurations, and mixed levels of resources for enterprise and edge domains. From either perspective of the data, the environment can be simplified to the information management of archived documents, document publications, and content knowledge.

By embracing a data perspective that combines relationships for unstructured knowledge representation with structured, document-centric relationships, the process of determining, modeling, and expressing relevance with semantic technologies can be performed. Our approach seeks to solve a combination of challenges within Information Extraction (IE), Information Management (IM), Semantic Information Modeling, Data to Information (D2I), and Quality of Service (QoS) Enabled Dissemination (QED).

\section{Modeling and Applying Directed Qualification for Analytics}
\label{sec:modeling}
Direct Qualification (DQ) for documents is the determination of their relevancy, considered either in combination or in isolation, and the appropriate semantic modeling of the results paired with the result of a stochastic analytic. DQ enables query capabilities for both the analytic results and the captured analytic qualification relationships for provenance. To formalize the problem faced when attempting to perform DQ we first need to explicitly define the concepts and steps involved. Primary concepts include qualification, provenance, specialization, relevancy scoring, occurrent and continuant state events, idempotent and non-idempotent analytics, and monotonicity.

The qualification of a relationship is an expression of how two entities, or in our case documents, are associated. For example, the foaf ontology supports properties that can declare whether two people know each other, but does not support the qualification of how, why, and to what degree they know each other. They also do not capture probabilistic relationships, such as the result of an analytic determining a strong likelihood that one person knows another. DQ provides the discoverability and traceability of the analytic qualification that resulted in the probabilistic relationship. Possible relationship qualifications that could be applied to the knows property of the foaf ontology are found in the Prov-O W3C Recommendation. The provided set of qualifications are general enough to provide generic use case solutions, but specific enough to extend with domain-based qualifiers if necessary. The Prov-O ontology, currently a W3C recommendation, seeks to provide a set of general provenance concepts and properties for interconnecting entities, activities, and agents. As an example, some supported properties are \texttt{wasAttributedTo}, \texttt{wasAssociatedWith}, \texttt{used}, \texttt{wasGeneratedBy}, \texttt{actedOnBehalfOf}, and \texttt{wasDerivedFrom}. Our work extends the provenance properties to support the expression of qualifications that are probabilistic in nature. We also extend the Prov-O concepts of entity, activity, and agent to relevancy counterparts such as idempotent, stochastic, or boolean analytics. These enhancements are intended to support modeling of analytic qualifications and results semantically.

Provenance supports relationship state changes, which have varying degrees of complexity via specialization. Semantic technologies serve multiple knowledge representation use cases, one being the expression of static relationships (Occurrent), the other being the expression of stateful (Continuant) relationships. Either one of these relationship types may be expressed independently as an RDF triple or message and document dependent as an RDF quad, using the source message / document as a possessive named graph. RDF and OWL are ideally suited to stateless, independent facts, as long as trust determinations, document traceability, and sourcing are non-critical features. However, in most real world use cases these features become critical to determining truth and relevancy. For example, if semantic relationships are extracted from multiple sources that disagree on a date, identity, event, target, mission, or other relevant fact, and these facts are expressed utilizing solely triples, a determination of truth must be made on which relationship is correct. Determining truth among conflicted relationships is nearly impossible without first determining trust levels of the extraction sources. Once a source becomes more trusted than another, a form of provenance is pulled back into the process, whether implicitly or explicitly.

The type of relationships are also important to consider. Some attributes of an asset may be occurrent (e.g., name, identity, asset type, etc.), while others are continuant (e.g., fuel level, latitude, longitude, role, etc.). Semantics, even when using instances of an entity, treat all relationships as occurrent, although there is allowance for limiting their cardinality. OWL, SPARQL, and most ontologies do not have a built in mechanism to support the distinction between occurrent and continuant relationships. In order to retrieve changes of state for a data or object attribute of an instance, that attribute must be explicitly defined within an ontology or an additional, customized layer of abstraction.

DQ and a relevancy ontology must support complex and diverse analytics modeling. Some analytics result in outputs with probabilistic relational implications, while others result in boolean results. Some analytics can be aggregated into a normalized set, while others must be evaluated individually. Some are idempotent, always resulting in the same value, regardless of whether they were executed successively. A stochastic analytic results in a non-deterministic value for the state of a relationships. For example, calculating the VSM similarity of two documents  may result in a particular keyword being correlated by a value of 5.6. Is this important or relevant? It lacks much value unless normalized or racked against a threshold acting as a heuristic determination.

DQ required the construction of supporting infrastructure, including a quad-store, a raw document storage DB, and a graph-based analytics framework. In our experiments we used both Parliament and Virtuoso as quad-stores (wrapping the Jena graph interface), flat files for raw document storage (referenced by a key-value pair hash map), and Jung for graph-based analytics. Each document is viewed concurrently as an independent publication and a semantic named graph, while the query results are autonomously mapped from a semantic graph result set to a traditional vertexes/edges graph tailored for analytics scoring.

Following the establishment of the infrastructure, appropriate semantic models for object instances need to be adopted, which supports semantic distinctions from assets and their state, as well as provides state traceability queries. Some semantic models adopt a constantly "present" based view that updates the instance of the mission, target, person, etc. with relationships reflecting any changes in its state. This is unwise because the state change's value can never be considered truly distinct from its identity URI. Specialization qualifications within the Prov-O ontology reduce relationship duplication while creating the capabilities for more advanced analytics. Jung, an open-source graph-based analytics framework, was extended to enable its built in analytics over semantic graphs represented using the Jena framework interfaces. Once the semantic and non-semantic graphs were made interoperable we constructed an experimentation harness that performed sample SPARQL queries and relevancy analytics over the test scenario. The initial set of analytics applied included "Betweenness," "PageRank," and "HITS."

DQ may not be necessary for all cases where analytics provenance is desired. Applicability can be determined by applying the test matrix in Figure \ref{fig:dq_decision_matrix} to a possible use case. In the case where a semantic data or object relationship is expressing an event that is occurrent and non-probabilistic in nature, DQ modeling will not yield any new insights over traditional semantic inferencing or reasoning. In the case where an entity-entity relationship is continuant, or state-based, but non-probabilistic in nature, the specialization provenance feature enables traceability for state management. DQ is applicable for all analytics-based value outputs, although it is important to note that DQ is implemented differently depending upon the nature of the event.

\begin{table}
\centering
\newcolumntype{C}[1]{>{\centering\let\newline\\\arraybackslash\hspace{0pt}}m{#1}}
\begin{tabular}{c c C{1.25in} C{1.25in}}

& & \multicolumn{2}{C{2.5in}}{State Change Event} \\
& & \multicolumn{1}{C{1.25in}}{Continuant} 		& \multicolumn{1}{C{1.25in}}{Occurrent} \\
\cline{3-4}
 \multirow{2}{*}[.8em]{\rotatebox[origin=c]{90}{Relationship Type}}	& \rotatebox[origin=c]{90}{~Analytic~}	& \multicolumn{1}{|C{1.25in}}{Specialization and Direct Qualification} & \multicolumn{1}{|C{1.25in}|}{Direct Qualification} \\
\cline{3-4}
 			& \rotatebox[origin=c]{90}{~Attribute~}	& \multicolumn{1}{|C{1.25in}}{Specialization}	& \multicolumn{1}{|C{1.25in}|}{Default Semantic Inferencing and Reasoning} \\
\cline{3-4}
\end{tabular}

\captionof{figure}{DQ Decision Matrix}
\label{fig:dq_decision_matrix}
\end{table}

\begin{figure*}[htb]
\centering
\begin{tikzpicture}
	
	\node[activity] (PRA) [label=above:{prov$:$Activity}] {PageRank\\Analytic};
	
	\node[attribute] (ra) [below = 1cm of PRA] {xsd:double};
	\node[attribute] (rt) [below left = 3cm of PRA] {xsd:double};
	\node[attribute] (rm) [below right = 3cm of PRA] {xsd:integer};
		
	\node[entity] (PRS) [label=above:{prov$:$Entity},above left = 3cm of PRA] {PageRank\\Score};
	
	\node[attribute] (rScore) [below left = of PRS] {xsd:double};
	\node[attribute] (dt) [below = of PRS, left = of PRA] {xsd:dateTime};

	\node[entity] (InfoA) [label=above:{prov$:$Entity},above right = 3cm of PRA] {Information A};
	
	\draw [conn] (PRS) -- (rScore) node [connlabel,midway] {relevancy:score};
	\draw [conn] (PRS) -- (dt) node [connlabel,midway,yshift=-.5cm] (foo) {prov:generatedAtTime};
	\draw [conn] (PRS) -- (PRA) node [connlabel,midway,yshift=3] {prov:wasGeneratedBy};
	
	\draw [conn] (PRA) -- (rt) node [connlabel,midway,xshift=-.5cm] {relevancy:tolerance};
	\draw [conn] (PRA) -- (ra) node [connlabel,midway] {relevancy:alpha};
	\draw [conn] (PRA) -- (rm) node [connlabel,midway,xshift=1cm] {relevancy:maxIterations};
	\draw [conn] (PRA) -- (InfoA) node [connlabel,midway,yshift=-4] {prov:used};
	
\end{tikzpicture}
\caption{Simple Entity to Entity DQ Model}
\label{fig:entity_to_entity_dq_model}
\end{figure*}
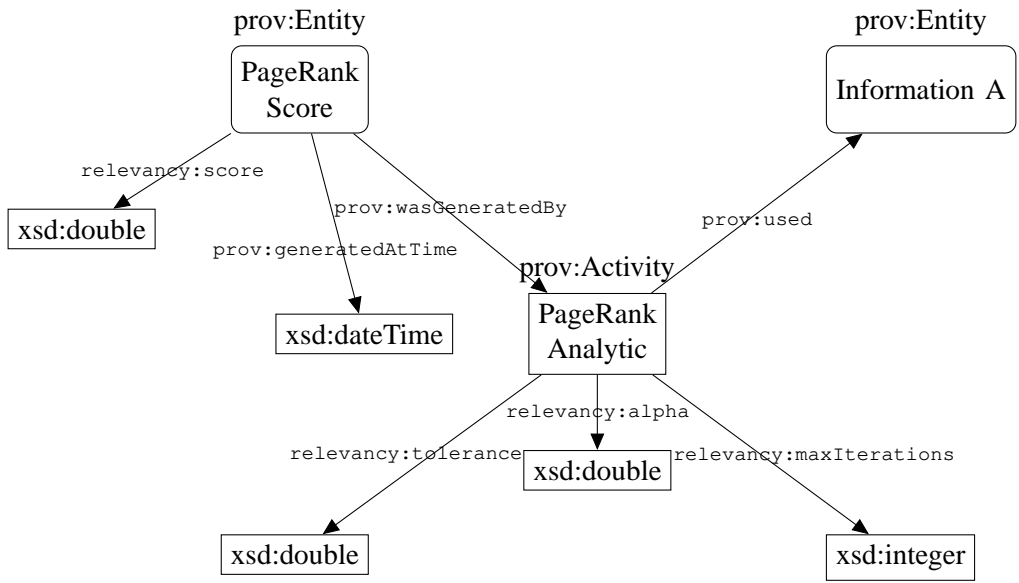

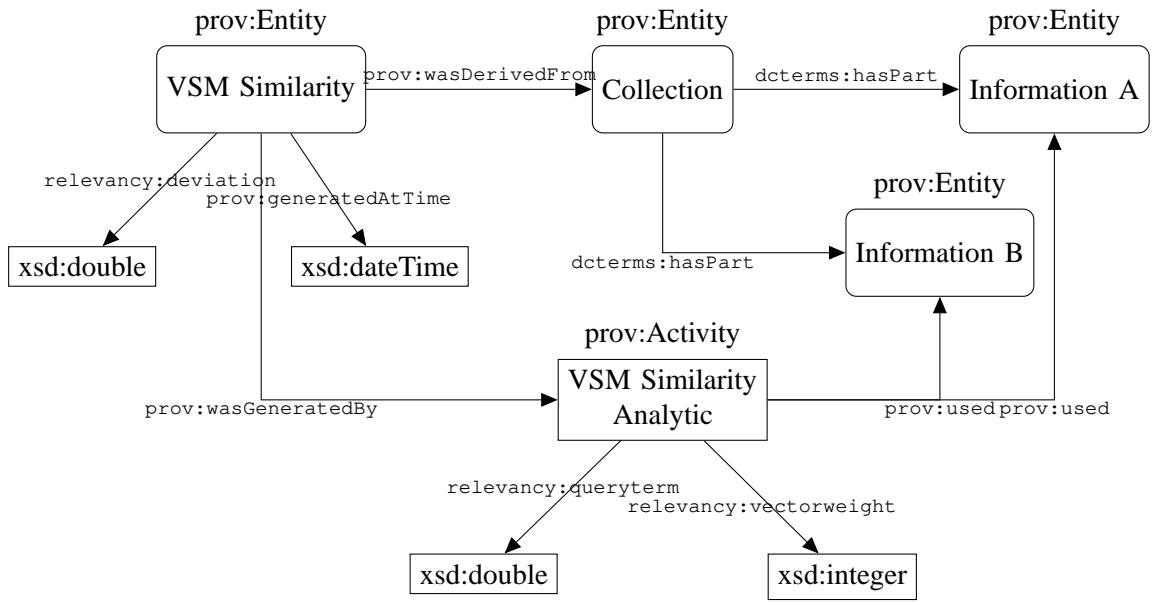
\begin{figure*}[htb]
\centering
\begin{tikzpicture}
	\node[entity] (VSMS) [label=above:{prov$:$Entity}] {VSM Similarity};
	\node[entity] (CL) [right = 3cm of VSMS,label=above:{prov$:$Entity}] {Collection};
	\node[entity] (IA) [right = 3cm of CL,label=above:{prov$:$Entity}] {Information A};
	\node[entity] (IB) [below left = 1cm and -1cm of IA,label=above:{prov$:$Entity}] {Information B};

	\node[activity] (VSMSA) [below = 3cm of CL,label=above:{prov$:$Activity}] {VSM Similarity\\Analytic};

	\node[attribute] (rqt) [below left = 1.5cm and 0cm of VSMSA] {xsd:double};
	\node[attribute] (rvw) [below right = 1.5cm and 0cm of VSMSA] {xsd:integer};

	\node[attribute] (rd) [below left = 1.5cm and 0cm of VSMS] {xsd:double};
	\node[attribute] (pg) [below right = 1.5cm and -1cm of VSMS] {xsd:dateTime};
	
	\draw[conn] (VSMS) -- (rd) node [connlabel,midway,yshift=5]{relevancy:deviation};
	\draw[conn] (VSMS) -- (pg) node [connlabel,midway,yshift=-5]{prov:generatedAtTime};\
	\draw[conn] (VSMS) |- (VSMSA) node [connlabel,midway,yshift=-5]{prov:wasGeneratedBy};

	\draw[conn] (VSMSA) -- (rqt) node [connlabel,midway,yshift=5]{relevancy:queryterm};
	\draw[conn] (VSMSA) -- (rvw) node [connlabel,midway,yshift=-5]{relevancy:vectorweight};

	\draw[conn] (VSMSA.east) -| (IB.south) node [connlabel,midway,yshift=-5]{prov:used};
	\draw[conn] (VSMSA.east) -| (IA.south) node [connlabel,midway,yshift=-5]{prov:used};

	\draw[conn] (VSMS) -- (CL) node [connlabel,midway,yshift=8]{prov:wasDerivedFrom};
	\draw[conn] (CL) -- (IA) node [connlabel,midway,yshift=8]{dcterms:hasPart};
	\draw[conn] (CL) |- (IB) node [connlabel,midway,yshift=-5]{dcterms:hasPart};
\end{tikzpicture}
\caption{Complex Entity to Entities DQ Model}
\label{fig:entity_to_entities_dq_model}
\end{figure*}

Occurrent relationships present the simplest opportunity to apply DQ. They are relatively simple to model, will never require adjustment, and have a common form of expression regardless of tye type of analytic being applied. Figure \ref{fig:entity_to_entity_dq_model} illustrates the application of the PageRank analytic to a document. The PageRank analytic applied to semantic node graphs produces dynamically adjusting normalized scores because the publication of subsequent documents grows the degree of interconnections for the document. This should, theoretically, predict an increased likelihood of popularity. This analytic differs from others such as the Vector Space Modeling (VSM) estimation for a single document. Regardless of subsequent published relationships VSM will result in the same normalized values of keyword frequency. Even if VSM estimates are aggregated for an entire collection of documents, the score will not change based on alterations within the knowledge graph. These aggregate document analytics, however, leave the monotonic realm of entity-to-entity DQ models, and enters the realm of entity-to-entities DQ models.

There are two complex use cases for applying DQ. One use case is continuant relationships that derive from applied analytics. The other use case is for qualification relationships that are non-monotonic. Monotonicity can mean different things depending on whether it is applied in mathematical functions, logic, or semantics, but our use of the term describes the uni-directional, single object capabilities that limit semantic relationships. Figure \ref{fig:entity_to_entities_dq_model} illustrates a non-monotonic relationship applying DQ between multiple documents. This is a complex case because DQ is being used as an alternative to reification in qualifying an analytic to multiple sources or causes. Applying DQ in these forms to analytics value results in the capability to semantically express both the provenance, inputs, qualifications, and results of analytics. Together these can validate the quality and relevance of the information.

\section{Experiment Setup}
\label{sec:experiment}
The mission-based test scenario consisted of 230 messages published over a period of 10 minutes. The message types included USMTF formatted ATOs, Intel Reports, Battle Damage Assessments, Close Air Support Requests, F15, Predator, and MRAP CoT Blue Force Tracking messages, and JTAC Red Force Tracking status observations. The semantic relationships created were produced by means of the extraction framework we created in our previous ICCRTS research ~\cite{Bryant_2013_Making}, with added support for GeoSPARQL location extractions.

Simulated mission data is published and semantically expressed via a set of indexers/extractors. The result of semantic processing is a semantic document represented via RDF/OWL relating internal values for details involving times, locations, missions, targets, points of contact, etc. An example document from a Blue Force Tracking extraction is shown in Listing \ref{lst:blue_force_extraction}.

\begin{lstlisting}[language=ttl,float=*,basicstyle=\small,breaklines=true,label=lst:blue_force_extraction,caption={Semantic Document Resulting From Semantically Processing a Blue Force Track.}]
<http://phoenix.rl.af.mil/im.owl#Payload?id1ff6f625e02118959d2b54af2aff>
	rdf:type im:Payload ;
	im:formatId "cot" ;
	im:type "a-f-A-M-F-Q-a" .

scenario:Pred1
	im:publishedInformation
		<http://phoenix.rl.af.mil/im.owl#Information?idd761961c4eb4918385775832c8a4> ;
	im:hasLocation 
		<http://www.lanuv.nrw.de/osiris/geometries/adce33b2-24c6-469a-8aac-27860d51df0b> ;
	im:publisherId "Pred1" .

<http://phoenix.rl.af.mil/im.owl#Information?idd761961c4eb4918385775832c8a4>
	rdf:type im:Information ;
	im:hasPayload <http://phoenix.rl.af.mil/im.owl#Payload?id1ff6f625e02118959d2b54af2aff> ;
	im:hasPublisher scenario:Pred1 ;
	im:informationType "mil.af.rl.cot" .
	im:involvesMission “247A” .
	im:involvesAsset “Pred1” .
	im:taskStart
		[ rdf:type time:Instant ;
		time:inDateTime
			[ rdf:type time:DateTimeDescription ;
			time:day "---11"^^<http://www.w3.org/2001/XMLSchema#gDay> ;
			time:hour "19"^^<http://www.w3.org/2001/XMLSchema#nonNegativeInteger> ;
			time:minute "00"^^<http://www.w3.org/2001/XMLSchema#nonNegativeInteger> ;
			time:month "--02"^^<http://www.w3.org/2001/XMLSchema#gMonth> ;
			time:timeZone ""^^<http://phoenix.rl.af.mil/&tz-world;ZTZ> ;
			time:year "2012"^^<http://www.w3.org/2001/XMLSchema#gYear>
			]
		] ;

<http://www.lanuv.nrw.de/osiris/geometries/adce33b2-24c6-469a-8aac-27860d51df0b> 
	<http://www.opengis.net/ont/geosparql#asWKT> "<http://www.opengis.net/def/crs/OGC/1.3/CRS84> POINT (4.86035239692792 48.41661096320327)"^^<http://www.opengis.net/ont/sf#wktLiteral> .
\end{lstlisting}

Ontology support includes common solutions for time, geospatial (GeoSPARQL), common elements (U-Core SL), and custom ontologies for mission planning, information management, and relevancy. Format and XML type determination is performed in a pre-processing stage prior to semantic extraction, if applicable. The Aperture open source project was adopted to provide the majority of the format and type determination solution, although customization for DoD formats was required. After stages for pre-processing, format determination, type determination, and semantic extraction completes, the execution of the analytics and DQ are executed as part of the query process.

\section{Relevancy}
\label{sec:relevancy}
Representing the relevancy of documents measured via analytics that score them, either independently or in concert with other documents, presents multiple obstacles. The first challenge is within provenance modeling itself. Modeling document relevancy should result in a cohesive solution that uses a clean, simple, and straightforward query expression. Unfortunately, many of the provenance-based qualification relationships apply to use cases where entities, activities, or agents have been qualified either singularly or in pairs, but not as a composite. This greatly increases the complexity of the query expression. Secondly, the complexity of reification within semantic technologies makes it difficult to express the results of an analytic qualified relationship between entities with a cardinality of greater than one. These issues can be overcome in various ways, some of which are described below.

Analytics that can process a semantic graph and enable relevancy determination must be applied selectively through best fit evaluations. For example, some analytic frameworks, such as Google's MapReduce or Natural Language Processing (NLP), operate over raw documents upon either their publication or being returned as a DB query result. These frameworks are tailored for document or key/value pair stores such as Hadoop or Cassandra. Semantic inferencing can create a form of analytics by applying an ontology relevant to the domain and setting up type, sameness, and equivalence rules. For example, if an extraction from two documents results in the determination that there are matching identifiers for a mission, target, asset, or person, then the respective attribute can be inferred to have ``sameness,'' an instance equivalence. This is a common feature of establishing property chains within OWL 2. Semantic reasoners support rule-logic that can determine if the value of a relationship has reached a particular threshold, but do not support features for analytics for graph traversal, analytics modeling, or state change events, which is where our approach fills the gaps.

Relevancy determination can supply critical C2 features for both general and tailored solutions for information quality, prioritization, and stochastic relationship models. Some forms of relevancy can be compared to enhancing a search engine. Documents found by keyword searches have long been ordered by estimating relevance by applying some kind of popularity analytic. This has generally remained unobtainable for semantic queries because the use cases have either not been document-based or were unable to be processed by non rule-logic analytics within semantic data sets.

Many web-oriented popularity, similarity, and clustering analytics appear to be well suited for semantic datasets. Our framework adopted an initial set of analytics for use, including PageRank, HITS, and Betweenness. These stochastic solutions fit well as the nature of a document-oriented semantic dataset is identical in many ways to the nature of the Internet itself. The Internet is an HTML representation of multiple knowledge domains overlayed upon a set of segmented documents (with unique URLs), while a document-oriented semantic dataset is an RDF/OWL representation of multiple knowledge domains overlayed upon a set of segmented named graphs (with unique URIs). Both datasets contain interlinkings between documents, the only difference being that the links on the Internet are not explicitly tied to a particular ontology.

\begin{figure*}[htb]
\centering
\includegraphics[scale=.45]{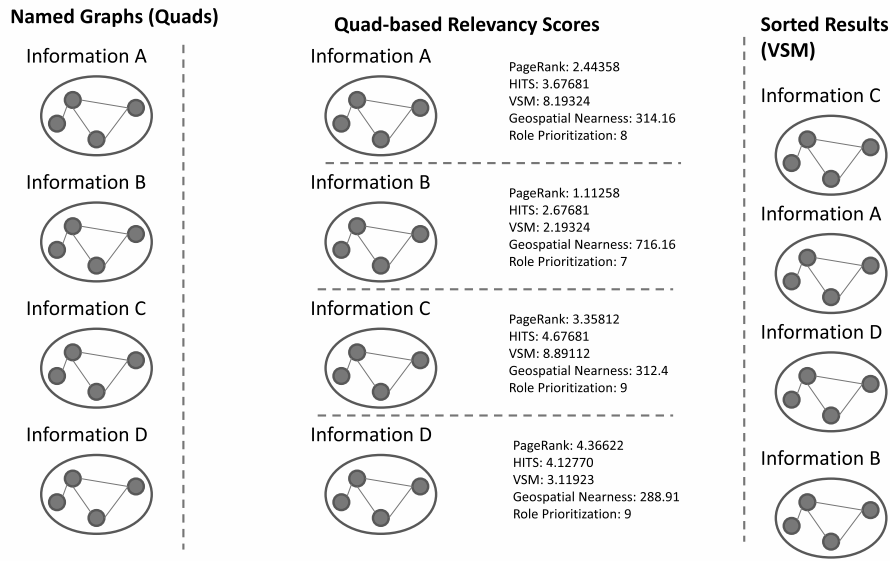}
\caption{Example Results of Applying Analytics with DQ}
\label{fig:analytic_results}
\end{figure*}

Our proof of concept for DQ seeks to apply relevancy analytics intended for the world wide web for semantic analysis. Our rsearch does not validate the quality of these analytics, but only seeks to model them appropriately via semantics. PageRank attempts to determine relevancy based upon links between webpages or semantic named graphs, using weight normalization to determine the likelihood of being at the node while doing random graph traversals. HITS seeks to determine the degree our named graphs are mutually reinforcing, and which named graph documents are authorities (highly interconnected). The output shown in Figure \ref{fig:analytic_results} is an example of DQ results for analytics applied to semantic named graphs, illustrating multiple analytic values for a query result set, and subsequently published to the semantic quad-store.

\section{Conclusion}
\label{sec:conclusion}
DQ models enable users to take advantage of multiple analytic paradigms and increase the quantity and quality of information available to users and administrators within a C2 domain. Semantic technologies can reap the benefits of graph-based analytics, traditionally considered separate and distinct, while allowing for the document-oriented nature of a traditional database and its paired features. These capabilities are obtained without degrading the semantic knowledge representation, but instead amplifies the capabilities so that SPARQL queries could be paired with additional document-based analytics available from other query languages and platforms.

Semantic modeling and persistence of the document DQ results enable some groundbreaking and novel features that should be more fully explored. One novel consequence is that SPARQL queries can order their semantic result set of document associated named graphs according to a set of prescribed analytics. For example, while keyword-based search analytics are commonly implemented by Google and other search providers in a way that orders results by a set of popularity and analytic oriented measures, semantic queries never have. By constraining named graphs within the quad-store to act as document references, DQ empowers the \texttt{ORDER BY} command within SPARQL to construct a document result set graph that can sort results by any expressed, normalized analytic score.

Another advantage of DQ is that analytic provenance and results can themselves be inference and reasoned upon. This enables an additional level of functionality that could reason a particular threshold value of an analytics score has been reached, triggering additional relationships with impacts affecting document prioritization, access control, security classification, or quality assessment.

These enhancements enable determinations of efficiency for different analytics, and have the potential to combine analytic-based queries with semantic queries. Applying these models, ontologies and approaches to an operational set of mission data can make that information, and its relevancy score results, more discoverable and of higher quality.

\IEEEtriggeratref{10}

\bibliographystyle{IEEEtran}
\bibliography{IEEEabrv,bibliography.bib}

\begin{thebibliography}{10}
\providecommand{\url}[1]{#1}
\csname url@samestyle\endcsname
\providecommand{\newblock}{\relax}
\providecommand{\bibinfo}[2]{#2}
\providecommand{\BIBentrySTDinterwordspacing}{\spaceskip=0pt\relax}
\providecommand{\BIBentryALTinterwordstretchfactor}{4}
\providecommand{\BIBentryALTinterwordspacing}{\spaceskip=\fontdimen2\font plus
\BIBentryALTinterwordstretchfactor\fontdimen3\font minus
  \fontdimen4\font\relax}
\providecommand{\BIBforeignlanguage}[2]{{%
\expandafter\ifx\csname l@#1\endcsname\relax
\typeout{** WARNING: IEEEtran.bst: No hyphenation pattern has been}%
\typeout{** loaded for the language `#1'. Using the pattern for}%
\typeout{** the default language instead.}%
\else
\language=\csname l@#1\endcsname
\fi
#2}}
\providecommand{\BIBdecl}{\relax}
\BIBdecl

\bibitem{Ding_2002_PageRank}
C.~Ding, X.~He, P.~Husbands, H.~Zha, and H.~D. Simon, ``Pagerank, hits and a
  unified framework for link analysis,'' in \emph{Proceedings of the 25th
  Annual International ACM SIGIR Conference on Research and Development in
  Information Retrieval}, ser. SIGIR '02.\hskip 1em plus 0.5em minus
  0.4em\relax New York, NY, USA: ACM, 2002, pp. 353--354.

\bibitem{Tous_2006_Vector}
R.~Tous and J.~Delgado, ``A vector space model for semantic similarity
  calculation and owl ontology alignment,'' in \emph{Proceedings of the 17th
  International Conference on Database and Expert Systems Applications}, ser.
  DEXA'06.\hskip 1em plus 0.5em minus 0.4em\relax Berlin, Heidelberg:
  Springer-Verlag, 2006, pp. 307--316.

\bibitem{Perry_2009_Geospatial}
M.~Perry, A.~Sheth, I.~B. Arpinar, F.~Hakimpour, and H.~A. Karimi, ``Geospatial
  and temporal semantic analytics,'' in \emph{Handbook of Research on
  Geoinformatics}, Hershey, PA, 2009, pp. 161--170.

\bibitem{Bryant_2013_Making}
J.~Bryant and M.~Paulini, ``Making semantic information work effectively for
  degraded environments,'' in \emph{Proceedings of the 18th International
  Command and Control Research and Technology Symposium}, June 2013.

\bibitem{NATO_2013}
N.~R.~T. Group, ``Executive overview of c2 agility,'' NATO Science and
  Technology Organization, Tech. Rep. STO-TR-SAS-085, November 2013.

\bibitem{DOD_2005}
U.~J.~F. Command, ``Command and control joint integrating concept,'' Department
  of Defense, Tech. Rep., September 2005.

\bibitem{Lebo_2013}
T.~Lebo, A.~Graves, and D.~L. McGuinness, ``Content\-preserving graphics,'' in
  \emph{Proceedings of Consuming Linked Data Conference}, 2013.

\bibitem{Bunescu_2004}
R.~Bunescu and R.~J. Mooney, ``Collective information extraction with
  relational markov networks,'' in \emph{Proceedings of the 42nd Annual Meeting
  on Association for Computational Linguistics}, ser. ACL '04.\hskip 1em plus
  0.5em minus 0.4em\relax Stroudsburg, PA, USA: Association for Computational
  Linguistics, 2004.

\bibitem{DOD_2010_Base-impl}
J.~Bryant, V.~Combs, J.~Hanna, G.~Hasseler, T.~Krokowski, B.~Lipa, J.~Reilly,
  and S.~Tucker, ``Phoenix base implementation: Final technical report,'' DTIC,
  Tech. Rep., 2010.

\bibitem{DOD_2010_Abstract}
J.~Bryant, V.~Combs, J.~Hanna, G.~Hasseler, R.~Hillman, B.~Lipa, J.~Reilly, and
  C.~Vincelette, ``Phoenix: An abstract architecture for information management
  final technical report,'' DTIC, Tech. Rep., 2010.

\bibitem{Barroso_2003}
L.~A. Barroso, J.~Dean, and U.~H{\"{o}}lzle, ``Web search for a planet: The
  google cluster architecture,'' \emph{IEEE Micro}, vol.~23, no.~2, pp. 22--28,
  March 2003.

\bibitem{Yakushiji_2006}
A.~Yakushiji, Y.~Miyao, T.~Ohta, Y.~Tateisi, and J.~Tsujii, ``Automatic
  construction of predicate-argument structure patterns for biomedical
  information extraction,'' in \emph{Proceedings of the 2006 Conference on
  Empirical Methods in Natural Language Processing}, ser. EMNLP '06.\hskip 1em
  plus 0.5em minus 0.4em\relax Stroudsburg, PA, USA: Association for
  Computational Linguistics, 2006, pp. 284--292.

\bibitem{Sudo_2003}
K.~Sudo, S.~Sekine, and R.~Grishman, ``An improved extraction pattern
  representation model for automatic ie pattern acquisition,'' in
  \emph{Proceedings of the 41st Annual Meeting on Association for Computational
  Linguistics - Volume 1}, ser. ACL '03.\hskip 1em plus 0.5em minus 0.4em\relax
  Stroudsburg, PA, USA: Association for Computational Linguistics, 2003, pp.
  224--231.

\bibitem{Hobbs_1997}
J.~R. Hobbs, D.~Applet, J.~Bear, D.~Israel, M.~Kameyama, M.~Stickel, and
  M.~Tyson, ``Fastus: A cascaded finite-state transducer for extracting
  information from natural-language text,'' in \emph{Finite-State Language
  Processing}.\hskip 1em plus 0.5em minus 0.4em\relax MIT Press, 1997, pp.
  383--406.

\bibitem{Schrage_2004}
M.~Schrage, ``The struggle to define agility,'' \emph{CIO Magazine}, August
  2004.

\bibitem{Shulstad_2011}
R.~A. Shulstad, ``Cursor on target: Inspiring innovation to revolutionize air
  force command and control,'' \emph{Air {\&} Space Power Journal}, vol. XXV,
  no.~4, pp. 19--28, Winter 2011.

\end{thebibliography}

\end{document}